\documentclass[12pt]{article}
\usepackage{jheppubmod,amsmath,amssymb}
\pdfoutput=1
\usepackage{color}
\usepackage{amsmath,amssymb}
\usepackage{comment}
\usepackage{braket}
\usepackage{mathtools}
\mathtoolsset{showonlyrefs}
\usepackage{psfrag}
\usepackage{array}
\usepackage{amssymb}
\usepackage{amsmath}
\usepackage{amsthm}
\usepackage{graphicx}
\usepackage{subcaption}
\usepackage{epstopdf}
	
\usepackage{color}
\usepackage{epsfig}
\usepackage[punctsep]{collref}
\collectsep[]{;}	

\def\pb[#1,#2]{\{#1, #2\}}
\def\deb[#1,#2]{[#1,#2]_{\text{D.B.}}}

\def\Or[#1]{{\text{O}}\left({#1}\right)}
\def\dotl[#1,#2]{\left\langle #1,\, #2 \right\rangle}
\def\dotlb[#1,#2]{\left\langle #1,\, #2 \right\rangle}
\def\dotlm[#1,#2]{\left[ #1,\, #2 \right]}
\def\dotp[#1,#2]{(\vect{#1} \cdot\vect{#2})}
\def\aff[#1,#2]{\hat{#1}(#2)}

\def\n4sym{{\cal N}=4 SYM}
\def\>{\rangle}
\def\<{\langle}
\def\weight[#1,#2,#3]{\{(#1),#2,#3\}}
\def\ads[#1]{$\text{AdS}_{#1}$}

\def\tarelr[#1]{\widetilde{a}^{\text{rel}}_{R#1}}
\def\Oright[#1]{{\cal O}_{R#1}}
\def\Oleft[#1]{{\cal O}_{L#1}}
\def\aleft[#1]{a_{L#1}}
\def\arelr[#1]{a^{\text{rel}}_{R#1}}

\def\tin{t_{\rm in}}
\def\tout{t_{\rm out}}
\newcommand{\tfdT}{|\Psi_T \rangle}

\newcommand{\tfd}{|{\Psi_{\rm tfd}}\rangle}
\newcommand{\tfdbra}{\langle {\Psi_{\rm tfd}|}}
\newcommand{\be}{\begin{equation}}
\newcommand{\ee}{\end{equation}}
\newcommand{\ba}{\begin{align}}
\newcommand{\ea}{\end{align}}
\newcommand{\bs}{\begin{split}}
\def\sess\end{split}

\newcommand{\vect}[1]{{#1}}

\def\tout{t_{\rm out}}

\author[a,b]{Rik van Breukelen}
\author[a,c]{and Kyriakos Papadodimas}
\emailAdd{rik.van.breukelen@cern.ch}
\emailAdd{kyriakos.papadodimas@cern.ch}
\affiliation[a]{Theoretical Physics Department, CERN, CH-1211 Geneva 23,
Switzerland}
\affiliation[b]{Geneva University, 24 quai Ernest-Ansermet, CH-1214 Geneva 4, Switzerland}
\affiliation[c]{Van Swinderen Institute for Particle Physics and Gravity, University of Groningen, Nijenborgh 4,
9747 AG Groningen, The Netherlands}

\abstract{Based on the work of Gao-Jafferis-Wall and Maldacena-Stanford-Yang, we observe that the time-shifted thermofield states of two entangled CFTs can be made traversable by an appropriate coupling of the two CFTs, or alternatively by the application of a modified quantum teleportation protocol. This provides evidence
for the smoothness of the horizon for a large class of entangled states related to the thermofield by time-translations. The smoothness of these states has some relevance for the firewall paradox and the proposal that some observables in quantum gravity may be state-dependent. We notice that quantum teleportation through these entangled states could be used in a laboratory setup to implement a ``time-machine'', which allows the observer
to travel  far in the future.
}

\preprint{\\\hspace*{\fill}
CERN-TH-2017-188
\vspace{-30pt}
}

\begin{document}
\title{Quantum teleportation through time-shifted AdS wormholes}

\maketitle

\newpage

\section{Introduction}

It has recently been demonstrated that the Einstein-Rosen bridge of the eternal AdS black hole can be made traversable via a particular double-trace deformation of the boundary CFTs \cite{Gao:2016bin}. Before the deformation, the two boundary theories were non-interacting and placed in a specific entangled state $\tfd = \sum_E {e^{-{\beta E \over 2}}\over \sqrt{Z}}|E,E\rangle$. The deformation creates shockwaves in the bulk, with negative average null energy,  which shrink the horizon of the black hole a little, allowing a particle to traverse the wormhole from one asymptotic region to the other. The deformation can also be formulated as a quantum teleportation protocol between the two CFTs \cite{Gao:2016bin,Maldacena:2017axo}. This setup has provided evidence for the smoothness of the horizon of the eternal black hole and for the ER=EPR proposal \cite{Maldacena:2013xja}.

In this paper we consider a similar experiment on a large class of states with different details in the entanglement between the CFTs. These states are of the form $\tfdT= e^{i H_R T} \tfd$, where $T$ is a parameter controlling the entanglement. 
While these states are as entangled as $\tfd$, 
they are {\it different} quantum states. We argue that the double-trace deformation (and the quantum teleportation protocol) can be modified to apply to each one of the states from this class. This provides evidence
that they all have a smooth horizon.

This simple observation has some interesting implications. First,  states of this family with $T>0$ can in principle be used in a lab setup to allow an observer crossing the wormhole to travel  far in the future in finite amount of proper time. During the
trip the observer is mostly in free fall. Second, for states with $T<0$, the bulk observer experiences evolution by finite proper time, while the elapsed time in the lab can become very small. Finally, the fact that we can establish the smoothness of this class of time-shifted states is of some interest for the firewall paradox \cite{Almheiri:2012rt, Almheiri:2013hfa, Marolf:2013dba} and the state-dependent proposal of \cite{Papadodimas:2012aq,Papadodimas:2013b,Papadodimas:2013,Papadodimas:2013kwa,Papadodimas:2015xma,Papadodimas:2015jra}.

We emphasize that the CFT correlators needed to support the claims of this paper are isomorphic to those relevant for \cite{Gao:2016bin, Maldacena:2017axo}. Hence, proving the traversability of the wormhole in the time-shifted states is  equivalent to the same proof for the TFD.

This paper is organized as follows: in section 2, we discuss the basic setup used by \cite{Gao:2016bin,Maldacena:2017axo}, which is the basis of the traversable wormhole. In section 3, we argue that time-shifted wormholes can be made traversable in a similar manner. In sections 4 and 5 we discuss this setup from a laboratory and a quantum-teleportation point of view. Finally, in section 6 we discuss some connections to the firewall paradox and state-dependence.

\section{\bf Traversable AdS wormholes}
In \cite{Maldacena:2001kr} it was proposed that two non-interacting copies of the same holographic CFT placed in the ``thermofield'' (TFD) entangled state
$$
\tfd = \sum_E {e^{-{\beta E \over 2}} \over \sqrt{Z}} |E\rangle \otimes |E\rangle
$$
are dual to the eternal black hole in AdS. This gravitational background can also be thought of as a wormhole connecting two asymptotic AdS regions. However, in this setup the wormhole is non-traversable, which is important for consistency
given that the boundary CFTs are non-interacting and hence no information can be exchanged between them.

\begin{figure}[t]
\centering
\includegraphics{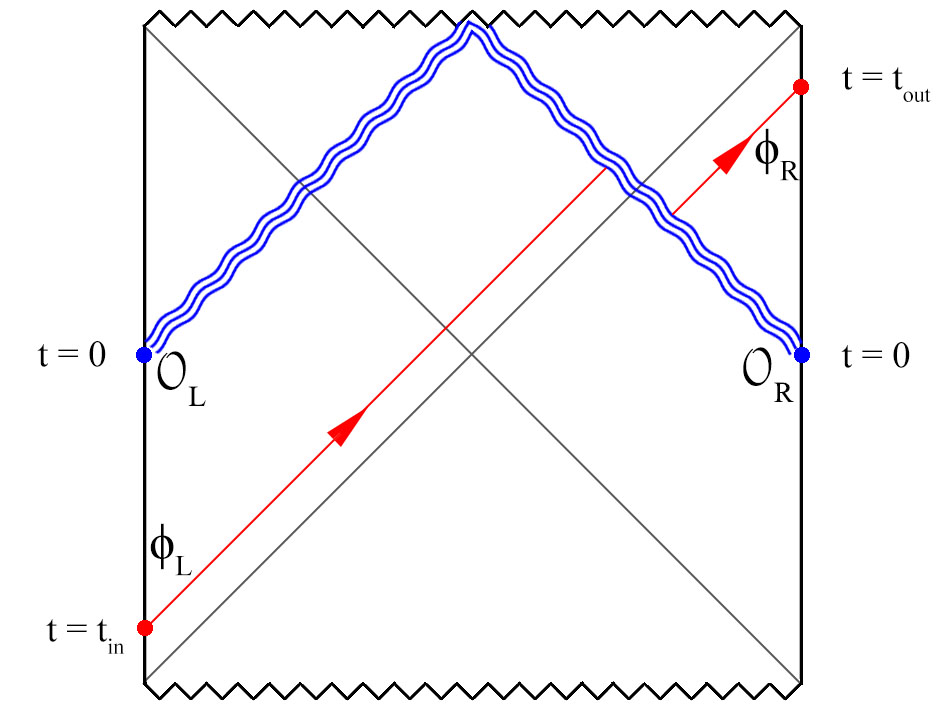}
\caption{A double trace deformation creates a shockwave which displaces the probe $\phi$, allowing it to escape from the black hole. The coordinates are discontinues at the shockwave, while the path of the probe is smooth.}
\label{fig1}
\end{figure}

It was realized in \cite{Gao:2016bin} that the wormhole can become traversable by coupling the two CFTs with a double-trace perturbation $e^{i g {\cal O}_L {\cal O}_R}$, which is turned on for a short time around $t=0$. Here ${\cal O}_{L/R}$ is a simple operator in the two corresponding CFTs.
By selecting the sign of $g$ appropriately, the perturbation creates negative null energy shockwaves falling into the black hole from both sides, see figure \ref{fig1}. This shrinks the horizon a little. As a result, an observer who dives from the left CFT at $t=\tin<0$ towards the black hole emerges on the right side and reaches close to the right boundary at $t=\tout>0$.  Here both $|\tin|$ and $\tout$ are of the order of the scrambling time $\beta \log S$. The details may depend on the theory and the form of the shockwave.

This setup is interesting because it allows us to probe the space-time in the interior of the wormhole purely in terms of 2-sided correlators of standard CFT operators. Directly probing the black hole interior from the CFT is more difficult, 
because we first have to define approximately local operators in AdS, which is non-trivial, especially behind the horizon. The setup of \cite{Gao:2016bin} bypasses the need to define these local bulk operators, as it probes the interior indirectly.
The observer is created by a local CFT operator $\phi_L(\tin)$, the perturbation is generated by CFT operators ${\cal O}_L(0){\cal O}_R(0)$ and the outgoing observer is detected by a local CFT operator $\phi_R(\tout)$. So the question of what happens to an observer falling through this wormhole can be translated into a computation of correlators of local CFT operators, the analogue of an S-matrix element in AdS. These well-defined (though difficult to compute in practice\footnote{In \cite{Maldacena:2017axo} relevant correlators were computed in the SYK model.}) CFT correlators can in principle provide evidence for the smoothness of the horizon of the eternal black hole and of the proposal \cite{Maldacena:2001kr}.

In \cite{Gao:2016bin} it was pointed out that this protocol is related to quantum-teleportation, and in \cite{Maldacena:2017axo} this quantum-teleportation protocol was made more explicit. First the observer is placed in the left CFT at $t=\tin$. At $t=0$ the operator ${\cal O}_L$ is measured. Depending on the resulting eigenvalue $o_L$, the unitary $e^{i g o_L {\cal O}_R}$ is applied on the right CFT. Then the quantum state of the system at time $t=\tout$ contains the observer emerging from the black hole into the right asymptotic AdS region. See also \cite{Susskind:2017nto}.

\section{Time-shifted wormholes}

In this paper we will work under the assumption that the TFD state can be made traversable for a semi-classical observer, as argued in \cite{Gao:2016bin, Maldacena:2017axo}. Using this as our starting postulate, we point out that there is a large class of other states with similar behaviour. These are states of the form
\be
\label{tfdshift}
\tfdT \equiv e^{i H_R T} \tfd = \sum_E {e^{-{\beta E \over 2}} \over \sqrt{Z}} e^{i E T}|E\rangle \otimes |E\rangle
\ee
It is important to realize that these are different quantum states from $\tfd$, due to the energy-dependent phases. 

\begin{figure}[ht]
\centering\centering
    \begin{subfigure}[b]{.48\textwidth}
        \includegraphics[width=\textwidth]{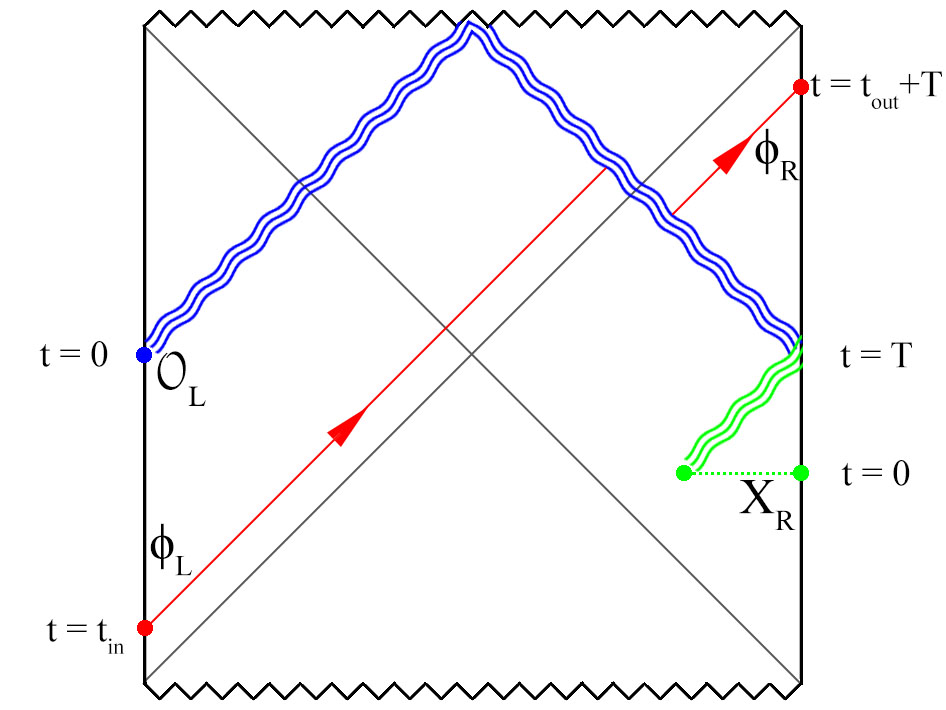}
        \caption{ }
        \label{fig2a}
    \end{subfigure}
    \begin{subfigure}[b]{.48\textwidth}
        \includegraphics[width=\textwidth]{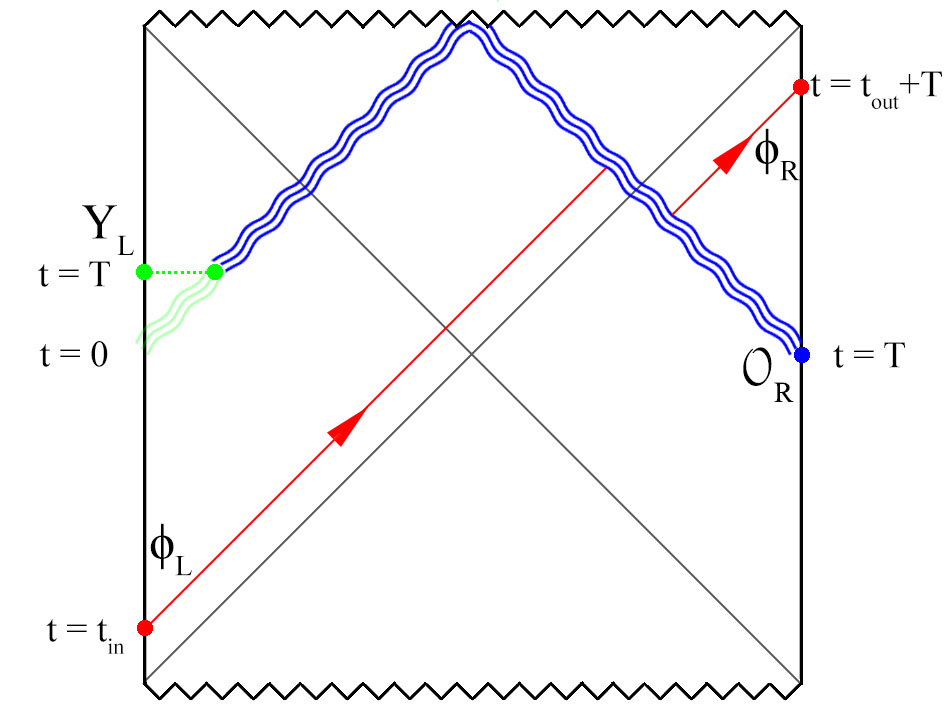}
        \caption{ }
        \label{fig2b}
    \end{subfigure}
\caption{a) In the time-shifted wormhole, with $T>0$, we need to act with a more complicated operator $X_R$ to receive the probe. b) Similar results can be achieved by using a precursor on the left CFT. Note that the Penrose diagrams can be misleading for precursors, because they may have a more involved bulk interpretation, see for example \cite{Heemskerk:2012mn}. However, the quantum state on the boundary after the end of the experiment can be reliably predicted.}
\label{fig2}
\end{figure}

The bulk interpretation of these time-shifted states, is that they are related to the usual eternal black hole by a large diffeomorphism, see for example \cite{Papadodimas:2015xma}. This is a diffeomorphism which acts as a time translation on the right boundary, but trivially on the left boundary. Since this is a large diffeomorphism (allowed by the boundary conditions), we are not supposed to mod-out by it. Instead, it maps a physical state to a different physical state. The states \eqref{tfdshift} can be represented as the usual eternal AdS black hole, but where the wormhole is ``anchored'' at different points in time on the two boundaries.
\vskip10pt
\noindent{\bf i) ${\bf T>0}$}
\vskip10pt
We argue that for every choice of $T>0$ the traversable wormhole protocol of \cite{Gao:2016bin} can be implemented: at $t=\tin<0$ an observer jumps into the left CFT. At $t=0$ we briefly couple the two CFTs by the operator $e^{i g {\cal O}_L(0) X_R(0)}$, where now $X_R(0)=e^{iH_RT}\mathcal{O}_R e^{-iH_RT}$ is the ``precursor'' of the operator ${\cal O}_R(T)$. Then at time $t=T + \tout$ the observer will come out in the right CFT, in exactly the same form as in the original experiment on the state $\tfd$. See figure \ref{fig2}. Alternatively, we could have used a precursor on the left, i.e. coupling the two CFTs with $e^{i g Y_L(T) \mathcal{O}_R(T)}$ where $Y_L(0)=e^{-iH_LT}\mathcal{O}_L e^{iH_LT}$, or some combination of left and right precursors at time $t$ satisfying $\tin < t<T$. The details of how the result of this experiment is isomorphic 
to that of the TFD is explained in appendix \ref{AppCTW}.

We emphasize that this statement is {\it exact}, even if $T$ is appreciably large. In other words, provided we accept that the protocol leads to a smooth traversable wormhole for the observer falling into the TFD, then same  can happen for all the other states, without any approximation. By tuning $T$ we can arrange that the observer will emerge significantly later in the future. Moreover, the quantum state of the observer after emerging in the right CFT will be exactly the same --- simply displaced in time. In particular her memories, and the amount of proper time that she will think has elapsed, will be the same\footnote{If we consider a big black hole in AdS whose radius $R_{\rm bh} \sim R_{\rm AdS}$ then the elapsed proper time will also be of order $R_{\rm AdS}$.} and independent of $T$. 
\begin{figure}
    \centering
    \begin{subfigure}[b]{.48\textwidth}
        \includegraphics[width=\textwidth]{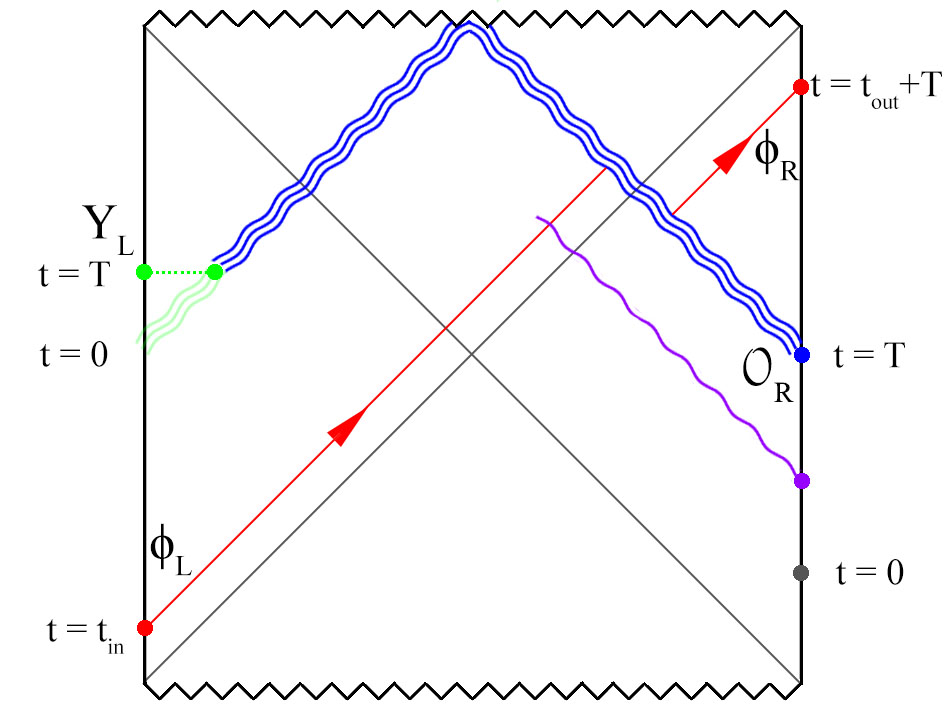}
        \caption{ }
        \label{fig3a}
    \end{subfigure}
    \begin{subfigure}[b]{.48\textwidth}
        \includegraphics[width=\textwidth]{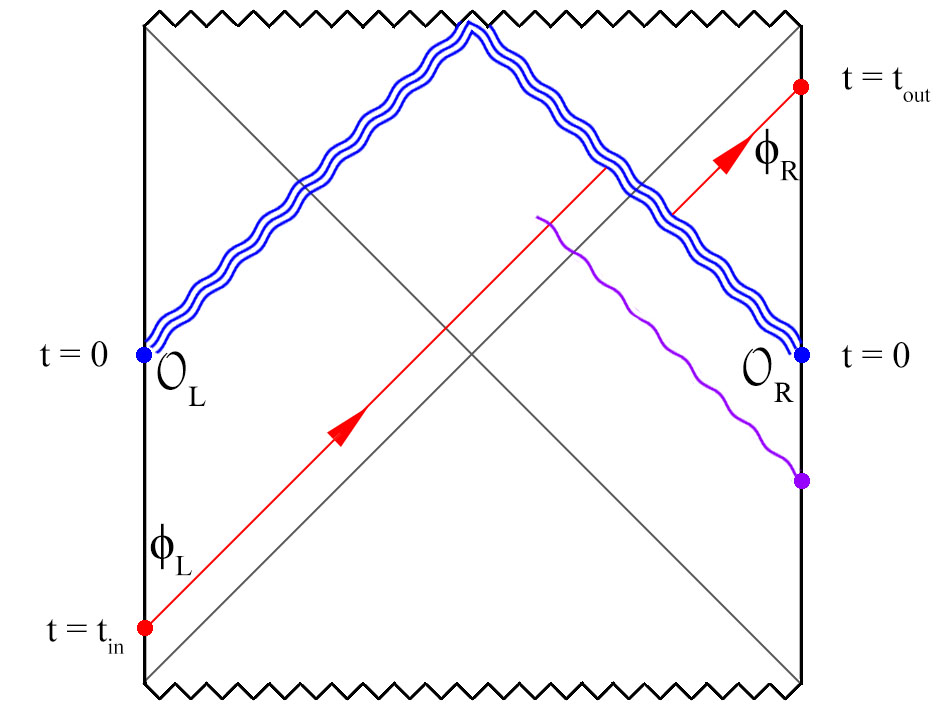}
        \caption{ }
        \label{fig3b}
    \end{subfigure}
    \caption{a) The memory of the probe can be modified by sending an early perturbation from the right. b) The same setup for the thermofield state.}
\label{fig3}
\end{figure}

While we can show that the observer emerges in the right CFT with memories of a smooth crossing of the wormhole, there is a logical possibility that the following scenario took place: during the crossing the observer actually experienced some unpleasant parts --- for example a firewall --- which killed her upon impact. Then the dynamics of the system ``resurrected`` the observer on the right side and in a state with memories corresponding to a smooth crossing. This scenario may sound  un-natural, but in some sense it is not so difficult to realize mathematically: for instance imagine an observer living inside a quantum system and that we act at $t=t_0$ with a unitary $U$ which kills him. At a later time $t_1$ we act with the precursor $e^{-i H (t_1-t_0)} U^{-1} e^{i H (t_1-t_0)}$. Then  the quantum state of the system for $t>t_1$ is the same as what it would have been, had we not acted with the first unitary which killed the observer. In this sense the sequence of $U$ at $t_0$ and its inverse precursor at $t=t_1$ kills and resurrects the observer. Moreover, when resurrected the observer has no memories of the fact that he had been killed. Notice that the unitary $U$ and its inverse precursor do not have to be fine-tuned with respect to the initial state of the observer in order to be able to resurrect him. 

In our setup the meaning of this question is whether 2-sided CFT correlators can exclude the possibility that the observer was killed when falling into the black hole from the left and resurrected when emerging in the right CFT.  

In order to directly address this question we would have to define local bulk observables which would be able to tell us what really happened in the middle of the bulk spacetime. As an easier alternative, we can send early signals from the right to probe the path of the observer, see figure \ref{fig3}.  These signals must be sufficiently weak, to avoid killing the observer or pushing the observer \cite{Shenker:2013yza, Shenker:2014} outside the window in which the coupling between the CFTs allows the extraction of the observer. We can then study if these signals modify the final quantum state of the outgoing observer in the way which is expected from effective field theory. If they do so, then we get additional evidence, though not definitive proof, that
nothing dramatic happened to the observer while crossing. Then this becomes again a statement of CFT correlators, which could in principle be computed.

We can see that these CFT correlators in the time-shifted TFD states are again isomorphic to the same correlators in the TFD state, provided that the signals from the right are sent with the appropriate time-shift. Hence if nothing strange happens to an observer crossing the TFD state as claimed by \cite{Gao:2016bin, Maldacena:2017axo}, then the same will be true for the time-shifted states. See appendix \ref{AppCTW} for some details.

\vskip10pt
\noindent{\bf ii) ${\bf T<0}$}
\vskip10pt

\begin{figure}
    \centering
    \begin{subfigure}[b]{.48\textwidth}
        \includegraphics[width=\textwidth]{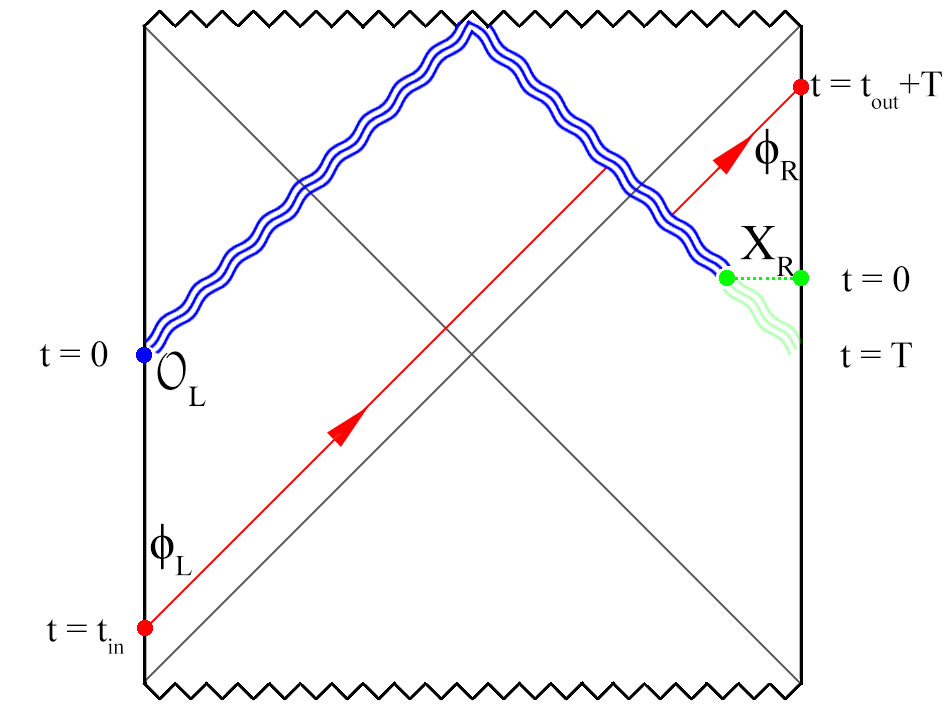}
        \caption{ }
        \label{fig4a}
    \end{subfigure}
    \begin{subfigure}[b]{.48\textwidth}
        \includegraphics[width=\textwidth]{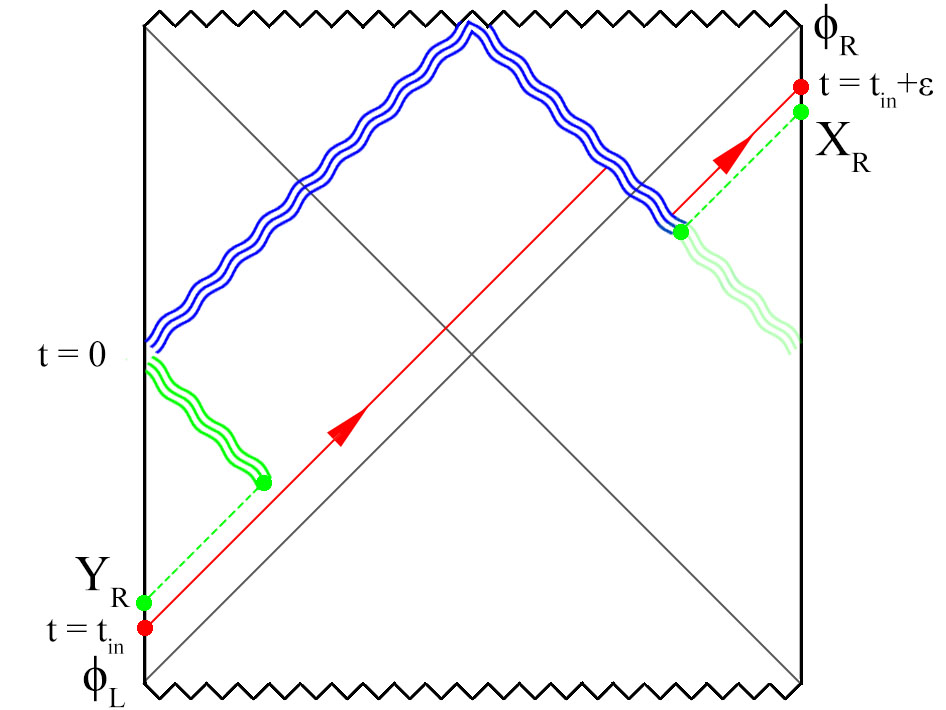}
        \caption{ }
        \label{fig4b}
    \end{subfigure}
    \caption{a) In the time-shifted wormhole, with $T < 0$, we can still recover the probe provided
that T is not to large. b) The extreme case in which we receive the probe almost immediately.}
\label{fig4}
\end{figure}

We can also consider states with $T<0$, with $|T|<\tout$. We can couple the two CFTs with the operator $e^{i g {\cal O}_L(0) X_R(0)}$, where again $X_R$ is the precursor of the operator ${\cal O}_R(T)$. Provided that $|T|<\tout$, the observer will emerge in the right CFT at $t=T+\tout>0$. Notice that in this setup the total lab time it takes for the observer to cross the wormhole is $T+ |\tin| +\tout$, which is less than $|\tin|+\tout$. So the crossing of the observer is accelerated for the lab frame, even though the proper time according to the observer is exactly the same.

Actually, we can shorten the lab time even more as follows. We throw the observer from the left at $t=\tin$ and we couple the two CFTs at $t=\tin + \epsilon$ by the operator $e^{i g Y_L X_R}$, where both are precursors. Then the observer comes out at $t= T+\tout$. Causality requires that $T+\tout > \tin + \epsilon$. This means we can push $T$ towards the negative values all the way to $T_{\rm min}= -|\tin|-\tout + \epsilon$. In this case the observer emerges on the right CFT almost immediately, even though according to her own experience the same (finite) amount of proper time as before has elapsed.

Notice that the full bulk interpretation of these protocols may be complicated due to the use of the precursors, which are complicated non-local operators. On the other hand we can reliably predict the exact quantum state of the observer --- with memories of smooth crossing and finite elapsed proper time --- as she emerges on the right CFT.

\section{Laboratory point of view}

In order to avoid possible confusions regarding the meaning of the time-shift in \eqref{tfdshift}, it is useful to think of the experiment in the following way.
We imagine a laboratory in a universe where gravity does not play an important role. We have two identical CFTs realized in some material in the laboratory. These are supposed to be holographic CFTs  dual to gravity in AdS.
There is only one common time, the laboratory time $t$. Each of the CFTs evolve with its own Hamiltonian, but since the CFTs live in the laboratory we identify the CFT time with the lab time $t_L=t_R = t$. 
The CFTs are prepared to be in a specific entangled state \eqref{tfdshift}. Hence it is more appropriate to think of the parameter $T$ as a ``dial'' that selects the initial state, rather than a time-shift. Of course preparing two CFTs in the TFD state or in one of its time-shifted cousins would be very difficult in practice, but possible in principle.

The laboratory technician can prepare a protocol where the observer is first injected into the left CFT at $t=\tin$. At $t=0$ the lab technician couples the two CFTs by the operator mentioned previously. Then at $t=T+\tout$ the observer emerges in the right CFT. From the point of view of the observer only a finite proper time elapses which is independent of $T$, but from the point of view of the lab the elapsed time is $T + |\tin|+\tout$, where $T$ can be arbitrarily large. Moreover throughout this experiment the observer is in free-fall, except for the (mild) interaction with the shockwave. We emphasize that the subjective experience of the observer is independent of the value of $T$ and in particular the strengh of the interaction with the schockwave is also independent of $T$.

It is interesting to notice that from the boundary point of view, the quantum information of the observer jumps from the left to the right CFT at $t=0$ when we couple the two CFTs. Then it stays scrambled in the right CFT for a long time, until it emerges in simple form at $t=T+\tout$. For instance, suppose that the observer on the left CFT carries a spin which is maximally entangled with some external reference spin. For $t<0$ the purification of the reference spin is in the left CFT. Right after $t>0$ the purification is in the right CFT but in scrambled form. Eventually at $T+\tout$ the purification of the reference spin is in the right CFT in terms of a simple spin carried by the observer.

\section{Quantum teleportation}

The  double-trace perturbation introduced in \cite{Gao:2016bin} can be slightly modified to be interpreted as a quantum teleportation protocol. In \cite{Maldacena:2017axo} this was described as follows: 
we make use of the fact that anything we do on the left boundary after acting with the double trace perturbation cannot affect the right boundary. For example, we could measure $\mathcal{O}_L$ just after the perturbation. Because $\mathcal{O}_L$ and $e^{ig\mathcal{O}_L\mathcal{O}_R}$ commute it would be equivalent to 
measure $\mathcal{O}_L$ just before the perturbation and then perturb  by $e^{i o_L {\cal O}_R}$, where $o_L$ is the eigenvalue measured.  Therefore, instead of acting with the double trace perturbation, the lab technician can implement the following protocol. First he releases the probe at $t=\tin$ in the left CFT. Then he measures ${\cal O}_L$ at $t=0$ and project onto one if its eigenstates with resulting eigenvalue $o_L$. Then he 
acts with a unitary $ e^{igo_L\mathcal{O}_R}$ at $t=0$ on the right CFT. The right CFT density matrix at the end of the teleportation protocol will be the same as the one in the double trace protocol, while the one on the left boundary will be different. 
 Notice that in the step of recording $o_L$ and selecting accordingly the unitary on the right we have the transfer of {\it classical} information from left to right, which is a part of a quantum teleportation protocol. 

The quantum teleportation protocol can be immediately realized for the time-shifted states: the lab technician first measures ${\cal O}_L$ at time $t=0$. Using the resulting eigenvalue $o_L$ he  applies at $t=0$ the unitary $U=e^{i go_L X_R}$ on the right CFT. Here $X_R(t=0)$ is the complicated precursor corresponding to the simple operator ${\cal O}_R(t=T)$.  Finally
at time $t=T+\tout$ the density matrix of the right CFT will be the same as in the experiment on the TFD at time $t=\tout$. This protocol is possible in principle, but it requires the use of the complicated operator $X_R$.

In the case that $T>0$ the lab technician can avoid having to use a complicated precursor, by performing an alternative ``time-delayed quantum teleportation protocol''.
He releases the  probe in the left CFT at $t=\tin$ ii) then he projects onto an eigenstate of $\mathcal{O}_L$ at $t=0$, recording the eigenvalue $o_L$. Then he waits until $t=T$ and he acts with a simple unitary $U=e^{igo_L\mathcal{O}_R}$ at $t=T$. Finally he considers the right CFT density matrix at $t=\tout+T$. 
This protocol  has the advantage that we do not have to use complicated precursors. We notice that we cannot use this protocol when $T<0$, as we would need to apply a unitary before the measurement of ${\cal O}_L$.

\section{Comments on state-dependence and the firewall}
\label{statedep}

The firewall paradox can be understood in its most precise formulation in terms of {\it typical} pure states of a 1-sided black hole in AdS. The argument starts by assuming that typical pure states have a smooth interior. It is then assumed that there
should exist some fixed linear operators acting on the Hilbert space, which correspond to local semiclassical observables behind the horizon. It is then shown that, according to bulk effective field theory, these observables would have to obey an algebra which is inconsistent with the density of states in the CFT \cite{Almheiri:2013hfa, Marolf:2013dba}. For this argument to work, it is important that we demand a smooth interior for a large class of states, i.e. for typical states. If we only look at a small number of states, then the paradox becomes less
sharp. In \cite{Papadodimas:2012aq,Papadodimas:2013b,Papadodimas:2013,Papadodimas:2013kwa,Papadodimas:2015xma,Papadodimas:2015jra} it was proposed that the paradox in its strongest form, i.e. for typical states, can be resolved by allowing the interior operators to depend on the state.

The smoothness of TFD state, as demonstrated by \cite{Gao:2016bin, Maldacena:2017axo} does not disprove the firewall, as the TFD is one particular state, while the firewall paradox becomes relevant when we consider {\it many} states. However, in \cite{Papadodimas:2015xma} it was pointed out that a version of the firewall paradox can also be formulated if we consider the entire family of time-shifted TFD states $e^{i H_R T}\tfd$ for all $T\in {\mathbb R}$. It was shown in \cite{Papadodimas:2015xma} that if we demand smoothness for all of these states, then we run into a firewall-like paradox, unless we accept that the interior operators are state-dependent. 
The argument of  \cite{Papadodimas:2015xma} was based on the assumption of smoothness for all time-shifted TFD states. This seems very plausible from the bulk point of view, since they are all related to TFD by a large diffeomorphism. However, it would be more satisfying if there was more
direct evidence for the smoothness of the time-shifted TFD states.

\begin{figure}[t]
\centering
\includegraphics{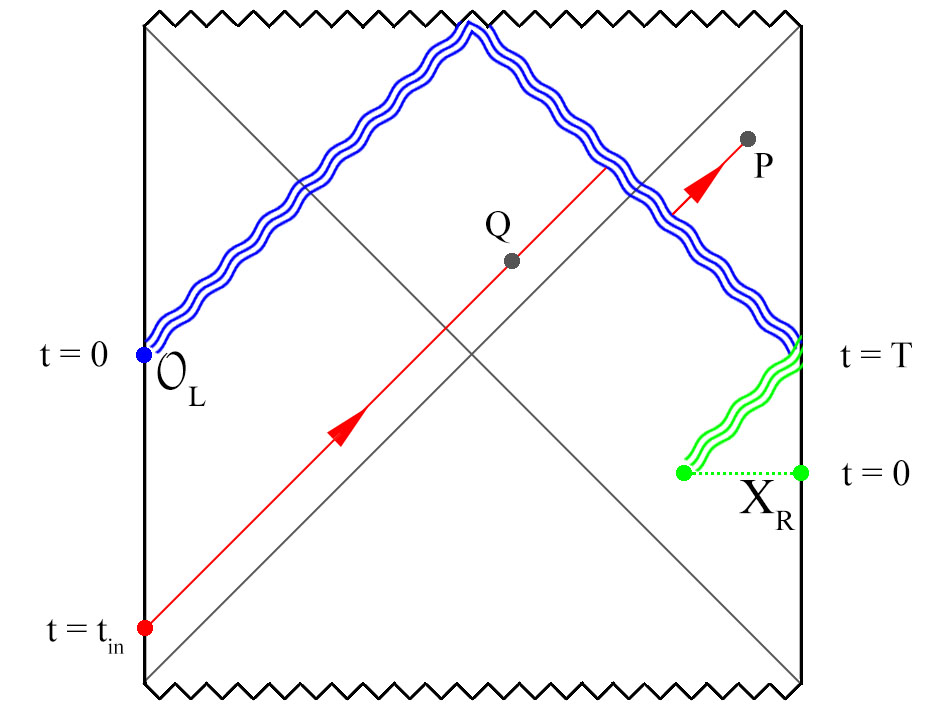}
\caption{Local operators at points $P,Q$ are state-dependent}
\label{fig5}
\end{figure}
In this paper we argued that by applying the teleportation protocol \cite{Gao:2016bin, Maldacena:2017axo} to the time-shifted states for all $T\in {\mathbb R}^+$, we find that all of them have a smooth interior. This disproves the firewall 
within a class of states where one would naively expect some firewall-like behavior\footnote{The argument of \cite{Papadodimas:2015xma} leads to a firewall-like paradox, even if we restrict to the family of states with $T>0$. To formulate this paradox we need to be able to take $T$ up to a time scale of order $e^S$.}. A natural explanation is that the interior operators in these
states are indeed state-dependent.

The class of time-shifted TFD states, together with the perturbation which allows the particle to escape the horizon, raise an interesting aspect of state-dependence for observables {\it outside the horizon}. Consider a local bulk operator at point $P$ in figure \ref{fig5}. According to the infalling observer,
this point is reached by diving in from the left CFT at $t_{\rm in}$ and freely-falling for a fixed amount of proper-time. For the infaller this relational prescription of the point $P$ is the same for all states, independent of $T$. However, the measurement of the operator at $P$ takes place at laboratory time $t=T+t_{\rm out}$. So this local operator at $P$ can be represented as {\it the same} operator in the Schr{\"o}dinger picture, however --- depending on the microstate --- it is applied by the infalling observer on the Schr{\"o}dinger-picture Hilbert space corresponding to a different time. 

We notice that the same property holds for local operators inside the horizon for this class of states, for example for a local operator at point $Q$. It is interesting to understand how this happens from the point of view of the infalling semiclassical observer, i.e.  how does she naturally identify the correct moment in time where the operators have to be applied.

\section{Discussion}

We investigated an extension of the traversable-wormhole protocol of \cite{Gao:2016bin, Maldacena:2017axo}, which has interesting physical interpretations.  We argued that using a larger class of entangled states, the time-shifted thermofield states, can lead to experiments involving time-travel in the lab.

General relativity allows time-travel to the future, by hovering near the horizon for a while and then flying away an observer can travel to the future. However, in order to move far in time, this method is not very pleasant, as it requires large proper accelerations. In this paper we described a more comfortable time-machine based on quantum entanglement. From the point of view of the observer the experience is pleasant, even if the desired time-difference is large. 

We notice that when the time shift $T$ becomes of the order of, or larger that the Poincare recurrence time, then the physical interpretation of the process must be done more carefully, since the observer may come out earlier than $T+\tout$, in the ``previous Poincare recurrence''

We also argued that for certain states with $T<0$ we can retrieve the observer almost immediately. One might worry that we can create a very fast computer by sending a computer through the wormhole, while there are fundamental bounds on computation speeds \cite{Lloyd:2000}. However, the CFTs creating the wormhole should be included as part of the computer, which will presumably respect the bounds.

It would be interesting to investigate the traversable wormhole protocol for more general entangled states of two CFTs. One particular class of such states would be superpositions of time-shifted thermofield states. Finally it would be interesting to investigate the possibility of traversing a single-sided black hole. In the case of the SYK model this was discussed in \cite{Kourkoulou:2017zaj}. Another class of candidate states in general holographic CFTs, which could be used as a starting point was proposed in \cite{Papadodimas:2017qit}. 


\begin{acknowledgments}
We would like to thank C. Bachas, J. de Boer, A. Gnecchi, M. Guica, D. Jafferis, A. Puhm, S. Raju, A. Stergiou, E. Verlinde for discussions and comments. R.v.B. was supported by NCCR SwissMAP of the Swiss National Science Foundation. K.P. would like to thank ENS, Paris and IHES  for hospitality during completion of this work and the Royal Netherlands Academy of Sciences (KNAW).

\end{acknowledgments}

\appendix
\section{Some details}
\label{AppCTW}
\subsection{Basic setup}
We recall the basic setup used in \cite{Gao:2016bin},\cite{Maldacena:2017axo}. We start with the TFD state, inject a probe (representing the observer) in the left CFT at time $t=\tin$ by acting with $e^{i\epsilon \phi_L}$. We couple the two CFTs with $e^{ig\mathcal{O}_L\mathcal{O}_R}$ for a very short time around $t=0$, and then calculate the expectation value of the outgoing probe on the right CFT with $\phi_R$ at $t=\tout$. All times are ``laboratory time''. We find it useful to think in terms of the wavefunction in the Schr{\"o}dinger picture as a function of the laboratoty time $t$. Ignoring the short amount of time that it takes to act with the operators mentioned above, the wavefunction is
\begin{align}
|\Psi(t)\rangle &= e^{-i (H_L +H_R) t}\tfd  & t<&\,\tin\\
|\Psi(t)\rangle &= e^{-i (H_L +H_R) (t-\tin)}e^{i\epsilon \phi_L} e^{-i (H_L +H_R) \tin}\tfd & \tin<\, t<&\,0\\
|\Psi(t)\rangle &= e^{-i (H_L +H_R)t} e^{ig\mathcal{O}_L\mathcal{O}_R} e^{i (H_L +H_R) \tin}e^{i\epsilon \phi_L} e^{-i (H_L +H_R) \tin}\tfd & 0<&\,t
\end{align}
Although the transitions are stated as sharp transitions, we should smear them a little to remove high energy modes.
Next we compute the expectation value of $\phi_R$ at time $t=\tout>0$:
\begin{align}
\label{finalcortfd}
\bra{\Psi(\tout)}\phi_R\ket{\Psi(\tout)}&=\tfdbra e^{i (H_L +H_R) \tin} e^{-i\epsilon \phi_L} e^{-i (H_L +H_R) \tin} e^{-ig\mathcal{O}_L\mathcal{O}_R}  e^{i (H_L +H_R)\tout} \\
&\times\phi_R e^{-i (H_L +H_R)\tout} e^{ig\mathcal{O}_L\mathcal{O}_R} e^{i (H_L +H_R) \tin}e^{i\epsilon \phi_L} e^{-i (H_L +H_R) \tin}\tfd \\
&=\tfdbra e^{-i\epsilon \phi_L(\tin)} e^{-ig\mathcal{O}_L(0)\mathcal{O}_R(0)}  \phi_R(t_{out}) e^{ig\mathcal{O}_L(0)\mathcal{O}_R(0)}  e^{i\epsilon \phi_L(\tin)} \tfd
\end{align}
Where we went from the Schr\"{o}dinger picture to the Heisenberg picture to make the times more clear. This  is the final correlator which can detect the excitation emerging in the right CFT after having traversed the wormhole.

In particular, by expanding the exponential $e^{i\epsilon\phi_L} \approx 1+i\epsilon\phi_L$, we obtain the commutator:
\begin{equation}
\braket{ [\phi_L(\tin),\phi_R(\tout)]}_V \neq 0
\end{equation}
Where $V$ denotes the double-trace pertubation. This was shown by \cite{Maldacena:2017axo} to be nonzero, thereby demonstrating information transfer between the boundaries. Moreover, it was argued that a multiparticle state is transfered without being destroyed, which is required for the path of the probe to be smooth. All this information is encoded in the correlator \eqref{finalcortfd}.

\subsection{Time-shifted states}
Let us now consider a time-shifted wormhole $\ket{\Psi_{T}} \equiv e^{i H_R T} \tfd$ with $T>0$. Note that, although we call this state a time-shifted state, we think of $T$ as a parameter controlling the entanglement of the state. We would like to perform the same steps as before. We start with the TFD state, inject a probe in the left CFT at time $t=\tin$ by acting with $e^{i\epsilon \phi_L}$. As we will see, in order to make this state traversalbe, we now have to couple the two CFTs with $e^{ig\mathcal{O}_L X_R}$ for a very short time around $t=0$, where $X_R$ is a {\it different} operator than ${\cal O}_R$. The quantum state is
\begin{align}
|\Psi_{T}(t)\rangle &= e^{-i (H_L +H_R) t} e^{i H_R T} \tfd  & t<&\,\tin\\
|\Psi_{T}(t)\rangle &= e^{-i (H_L +H_R) (t+t_w)}e^{i\epsilon \phi_L} e^{-i (H_L +H_R) \tin} e^{i H_R T} \tfd  & \tin<\, t<&\,0\\
|\Psi_{T}(t)\rangle &= e^{-i (H_L +H_R)t} e^{ig\mathcal{O}_LX_R} e^{-i (H_L +H_R) t_w}e^{i\epsilon \phi_L} e^{i (H_L +H_R) t_w} e^{i H_R T} \tfd  & 0<&\,t
\end{align}
We will now see that if we take $X_R$ to be the precursor of the operator $O_R(T)$ then the quantum state of the right CFT at $t=T+t_{out}$ will be {\it exactly} the same as that of the TFD state at $t=t_{out}$. In particular it will contain the particle emerging out of the black hole in exactly the same form as in the TFD. Note that, because the right Hamiltonian commutes with left operators, we can rewrite the coupling between the boundary systems as follows $ e^{ig\mathcal{O}_LX_R} = e^{ig\mathcal{O}_L e^{iH_RT}\mathcal{O}_R e^{-iH_RT}}= e^{iH_RT} e^{ig\mathcal{O}_L\mathcal{O}_R} e^{-iH_RT}$.
We can diagonse this by computing the expectation value of operator $\phi_R$ on this state.
\begin{align}
\bra{\Psi_{T}(\tout)}\phi_R\ket{\Psi_{T}(\tout)}&=\tfdbra  e^{-i H_R T_1} e^{i (H_L +H_R) \tin} e^{-i\epsilon \phi_L} e^{-i (H_L +H_R) \tin}\\
&\times e^{iH_RT_2} e^{-ig\mathcal{O}_L\mathcal{O}_R} e^{-iH_RT_2} e^{i (H_L +H_R)\tout} \phi_R e^{-i (H_L +H_R)\tout}\\
&\times e^{iH_RT_2} e^{ig\mathcal{O}_L\mathcal{O}_R}e^{-iH_RT_2} e^{i (H_L +H_R) \tin}e^{i\epsilon \phi_L} e^{-i (H_L +H_R) \tin} e^{i H_R T_1} \tfd \\
&=\tfdbra e^{-iH_R(T_1-T_2)} e^{-i\epsilon \phi_L(\tin)} e^{-ig\mathcal{O}_L(0)\mathcal{O}_R(0)}  \phi_R(\tout-T_2) \\
&\times e^{ig\mathcal{O}_L(0)\mathcal{O}_R(0)}  e^{i\epsilon \phi_L(\tin)} e^{iH_R(T_1-T_2)}\tfd
\end{align}
For the case that $T_1=T_2=T$ we obtain:
\begin{equation}
\bra{\Psi_{T}(\tout)}\phi_R\ket{\Psi_{T}(\tout)}_{V'} = \bra{\Psi_{0}(\tout-T)}\phi_R\ket{\Psi_{0}(\tout-T)}_V
\end{equation}

The time-shifted state is related to the normal state by a delay in the response. This is an {\it exact} statement: no approximations where made. The time-shifted state is just as traversable as the thermofield double state. However, the probe is received at a later time $t=\tout+T$. The probe does not feel this difference: the proper time of the proper does not depend on $T$. The time shifted wormhole can, therefore, be used as a time-machine. By tuning the state and the corresponding perturbation, we can shift the probe by an arbitrary amount to the future, without changing the perception of the probe. An alternative to acting with the precursor on the right CFT, is acting with a precursor on the left CFT $e^{igY_L\mathcal{O}_R}$, or acting with precursors on both sides $e^{igY_LX_R}$. These options show the same results. Moreover, we can extend these results for $T<0$ up to $T=-|\tin|-\tout+\epsilon$, where $\epsilon$ is some small number representing some limitations. For example, the sending of the probe, the scattering, and the receiving of the probe, are either smeared or nonlocal, and the precursors should not act within those areas. Therefore, we cannot receive the probe with zero time elapsed.

\subsection{Additional shockwaves}
It is necessary for the path of the probe to be smooth that the probe remembers what happened before encountering the shockwave, otherwise one may argue that the probe was killed earlier and regenerated by the shockwave. We consider the following state to study this:
\begin{align}
|\Psi_{T}(t)\rangle &= e^{-i (H_L +H_R) t} e^{i H_R T} \tfd  &t<\tin\\
|\Psi_{T}(t)\rangle &= e^{-i (H_L +H_R) (t-\tin)}e^{i\epsilon \phi_L} e^{-i (H_L +H_R) \tin} e^{i H_R T} \tfd  &\tin<\, t<t_s\\
|\Psi_{T}(t)\rangle &= e^{-i (H_L +H_R) (t-t_s)} e^{ia \varphi_R}e^{-i (H_L +H_R) (t_s-\tin)}e^{i\epsilon \phi_L} e^{-i (H_L +H_R) \tin} e^{i H_R T} \tfd  &t_s<\, t<0\\
|\Psi_{T}(t)\rangle &= e^{-i (H_L +H_R)t} e^{ig\mathcal{O}_LX_R}  e^{i (H_L +H_R) t_s} e^{ia \varphi_R}\\
&\times e^{-i (H_L +H_R) (t_s-\tin)}e^{i\epsilon \phi_L} e^{-i (H_L +H_R) \tin} e^{i H_R T} \tfd  &0<t
\end{align}
Where $e^{ia \varphi_R}$ generates a shockwave coming from the right, at some early time $t_s$. It is important that this shockwave is extremely weak, otherwise it will kill the probe. The double trace perturbation should still be able to extract the probe after the interaction of the additional shockwave. Moreover we want to use the same double trace perturbation, whether we would send this additional shockwave or not. An alternative would be to modify the double trace perturbation to remove this additional shockwave. That setup, however, cannot answer whether the memories of the probe are genuine. We rewrite the state, for $t>0$, in such a way that it is clear that it corresponds to the same experiment in the thermofield state. The correlator follows directly from that.
\begin{align}
|\Psi_{T}(t)\rangle &= e^{-i (H_L +H_R)t} e^{ig\mathcal{O}_LX_R}  e^{i (H_L +H_R) (t_s)} e^{ia \varphi_R}e^{-i (H_L +H_R) (t_s-\tin)}e^{i\epsilon \phi_L} e^{-i (H_L +H_R) \tin} e^{i H_R T} \tfd \\
&= e^{-i (H_L +H_R)(t-T)} e^{ig\mathcal{O}_L\mathcal{O}_R}  e^{i (H_L +H_R) (t_s-T)} e^{ia \varphi_R}e^{-i (H_L +H_R) ((t_s-T)-\tin)}e^{i\epsilon \phi_L} e^{-i (H_L +H_R) \tin} \tfd
\end{align}
If we calculate $\langle\phi_R\rangle$ in this state we get the same response as in the thermofield state, with both the response and the additional shockwave being shifted by $T$. Thus we could extract information about the additional shockwave from the response. 
\bibliographystyle{utphys}
\bibliography{references}
\end{document}